\documentclass[12pt]{iopart}

\newcommand{\eqref}[1]{(\ref{#1})}
\usepackage{graphicx}
\usepackage{iopams}

\begin{document}

\title{Condensate Formation in a Chiral Lattice Gas}

\author{Boyi Wang$^{1,2,3}$, Frank J\"ulicher$^{3,4,5}$ and Patrick Pietzonka$^{3,6}$\footnote{corresponding author}}

\address{%
$^1$Beijing National Laboratory for Condensed Matter Physics and Laboratory of Soft Matter Physics, Institute of Physics, Chinese Academy of Sciences, Beijing 100190, China}
\address{%
$^2$School of Physical Sciences, University of Chinese Academy of Sciences, Beijing 100049, China}
\address{%
$^3$Max Planck Institute for the Physics of Complex Systems, Dresden 01187, Germany}
\address{%
$^4$Center for Systems Biology Dresden, Pfotenhauerstrasse 108, 01307 Dresden, Germany}
\address{%
$^5$Cluster of Excellence Physics of Life, TU Dresden, 01062 Dresden, Germany}
\address{%
$^6$SUPA, School of Physics and Astronomy, University of Edinburgh, Peter Guthrie Tait Road, Edinburgh EH9 3FD, United Kingdom
}

\eads{\mailto{boyiw@pks.mpg.de}, \mailto{julicher@pks.mpg.de}, \mailto{p.pietzonka@ed.ac.uk}}

\vspace{10pt}
\begin{indented}
\item[]\today
\end{indented}
\date{\today}

\begin{abstract}
We investigate the formation of condensates in a binary lattice gas in the presence of chiral interactions. These interactions differ between a given microscopic configuration and its mirror image. We consider a two dimensional lattice gas with nearest-neighbour interactions to which we add interactions involving favoured local structures that are chiral. We focus on favoured local structures that have the shape of the letter L and  explore  condensate formation through simulations and analytical calculations. At low temperature, this model can exhibit four different phases that are characterised by different periodic tiling patterns, depending on the strength of interactions and the chemical potential. When particle numbers  are conserved, some of these phases can coexist. We analyse the structure and surface tension of interfaces between coexisting phases and determine the shapes of minimal free energy of crystalline condensates. We show that these shapes can be quadrilaterals or octagons of different orientation and symmetry.
\end{abstract}

%\maketitle

%\tableofcontents

\section{Introduction}

Chirality, which refers to asymmetry under mirror imaging, is common in many forms of matter. It is ingrained on the microscopic scale of various model systems, such as chiral granular gases \cite{tsai2005chiral}, chiral squirmers \cite{burada2022hydrodynamics}, spontaneously rotating droplets \cite{wang2021spontaneous}, L-shaped colloidal chiral microswimmers \cite{kummel2013circular}, or cholesteric
liquid crystals with self-organized helical superstructures \cite{bahadur1990liquid,zheng2016three}. In biological matter,
Chiral helical structures in actin filaments and microtubules cause motors moving along them to also rotate around them \cite{sase1997axial}, which leads to the formation of chiral patterns by cytoskeletal filaments \cite{tee2015cellular} and play a role in left-right symmetry breaking in biology \cite{furthauer2013active, naganathan2014active}.
On macroscopic scales, chiral matter can display unusual effects such as odd viscosity, elasticity, or viscoelasticity \cite{fruchart2023odd}, unique phase separation behaviours and edge currents \cite{zhao2021emergent,ding2024odd, soni2019odd}.

In this paper, we investigate the role of chirality in binary mixtures. In particular, we want to understand how chiral interactions between solvent and solute particles affect the physical properties of condensates on a macroscopic scale. For simplicity, we focus on planar-chirality in two dimensions, which means a 2D pattern cannot be superposed on its mirror image by any combination of rotations and translations~\cite{bruice2017organic}. Such a planar setting is similar to the adsorption of chiral molecules (or atoms that form such molecules) on a planar substrate~\cite{ortega2000extended}.

The classic way to study binary mixtures starts with a lattice gas model~\cite{huggins1941solutions,flory1942thermodynamics,rothman1994lattice,safran2018statistical}. Such a model describes particles of two types occupying sites on a square lattice with interactions between nearest-neighbor particles. It is equivalent to the Ising model, with particle occupancy corresponding to two opposite spin states.
With number conservation of particle types, it already captures the basic principles of phase separation. We aim to build on this classic model and break chiral symmetry in a minimal fashion. Since the nearest-neighbour interactions of the Ising model are inherently achiral, such a model needs to include interactions between next-nearest neighbours as well~\cite{domb1951order}. 

A simple way to introduce complex next-nearest-neighbour interactions in a lattice model is to define a so-called favoured local structure (FLS) \cite{ronceray2011variety}. Any local realisation of such a structure leads to the reduction of the total energy by an amount that is characteristic for the strength of the FLS interaction.

Previous studies of FLS interactions have focused on the order-disorder phase transition~\cite{ronceray2011variety,ronceray2012geometry,ronceray2013influence,ronceray2014multiple} and on disordered systems like spin glasses, where the interaction is akin to an FLS, but random~\cite{sherrington1975solvable,derrida1980random}. A complementary way to model chirality in a lattice gas is to consider nearest-neighbour interactions only, but to give particles a rotational degree of freedom that the interactions depend on~\cite{lombardo2009thermodynamic}. Our focus here is on the formation of condensates with FLSs far below any critical temperature. There, the system can be considered to be in the ground state that minimises the total energy. 

We first focus on the ground states without number conservation, where particles of the two types can be exchanged via reservoirs. We find that the preference of a chiral FLS can induce chiral tiling patterns across the lattice. Using analytical methods, we determine different ground states, depending on the strength of the FLS interaction and the exchange chemical potential. These results are complemented by Monte-Carlo simulations with Metropolis sampling at small temperature.

To investigate phase separation and coexistence, we consider a lattice with particle number conservation. This is the setting where we observe the formation of condensates as patches of one phase immersed in a second one. These condensates can range from liquid-like droplets where the Ising interaction dominates to crystalline structures induced by the FLS interaction. We analytically construct a phase diagram of scenarios of phase coexistence, changing with both number concentration and the strength of the FLS interaction. We find that breaking local chiral symmetry leads to the formation of condensates with chiral geometry. Additionally, we examine the interfacial behaviours of phase coexistence in finite systems and develop a method to define and calculate interfacial energy. This method allows us to determine the shape of the condensate in the ground state. These findings are complemented by Monte-Carlo simulations with Kawasaki dynamics.

\section{Lattice gas with a favoured local structure}

A simple model for a phase separating system in two dimensions is the Ising model. The two possible states of every lattice site may be interpreted as spin states, or, as we do here, as two different types of particles (A and B) in a lattice gas.
We consider a square lattice with $N$ sites and periodic boundary conditions, where each site $i=1,\dots, N$ can take the values $\sigma_i = +1, -1$, representing two components A and B. The total number of A is $N_A=\sum_i \delta_{\sigma_i,1}$, the total number of B is $N_B=\sum_i \delta_{\sigma_i,-1}$. 

\begin{figure}
  \includegraphics[width=1\textwidth]{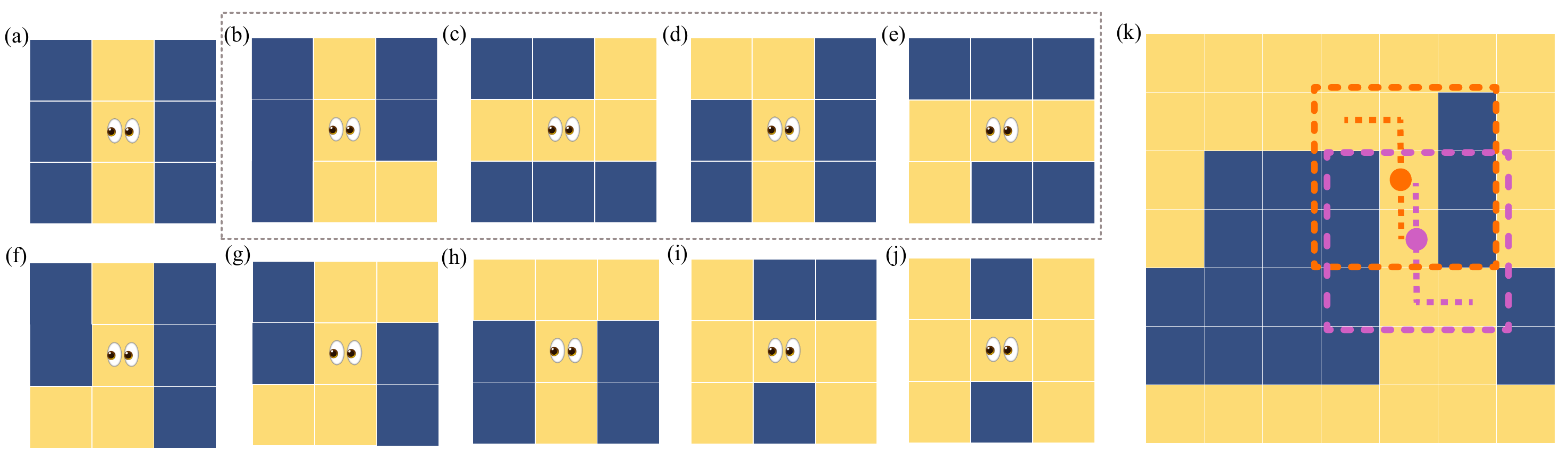}
   \caption{\label{fig:FLS}% 
   Examples for FLS defining a multi-body interaction centered around the lattice sited marked by the pair of eyes. Mirror images of a chiral pattern [such as the L shape (b) and its mirror image (f)] are treated as distinct FLS, while different rotations of the same pattern [(b)-(e)] contribute equally to the interaction. Panel (k) shows an example lattice configuration, for which we count $N_K=2$ instances of the L-shape FLS. Circular dots represent centres of each L, with the eight neighbouring cells and orientations of the Ls indicated as dashed lines.
   }
\end{figure}

We consider a nearest-neighbour interaction of strength $J>0$. This interaction favours equal states of a site and its neighbours, reflecting the basic and well understood behaviour of a solute (type A) that tends to phase separate from solvent particles (type B) below a critical temperature. 

Beyond the Ising interaction, FLSs allow us to consider multi-body interactions among the particles of type A~\cite{ronceray2011variety}. For our square lattice, this type of interaction takes into account the states of the $3\times3$ local neighbourhood around every central particle of type A. Every such neighbourhood that exactly matches a pre-defined pattern reduces the energy of the system by the value $K$, a parameter for the strength of this type of interaction. Fig.~\ref{fig:FLS} shows a small set of examples for such FLS, which we refer to by their resemblance to the letters I, L, J, S, T, h, and H. In order to give the model invariance under $90^\circ$ rotations, rotations of a pattern (such as of the L-shaped pattern in Fig.~\ref{fig:FLS}b-e) are treated as the same local structure. Crucially, however, mirror images of a chiral pattern, such as the mirror image Fig.~\ref{fig:FLS}f of the L shape, do not count as a realisation of the same FLS. Note that multi-body interactions involving only the four nearest neighbours of every lattice site would not allow for any chiral patterns, hence our focus on the $3\times 3$ neighbourhoods involving diagonal neighbours as well. Our choice for this type of multi-body interaction is primarily due to its simplicity, allowing us to study the role of chirality in condensate formation in a minimal way. Physical systems that could be modeled in this way are the adsorption of atoms to a surface where they form chiral molecules, or enzymes that bind to each other via allostery~\cite{herve1989allosteric}, i.e., with binding energies that depend on the state of other binding sites.

For every microstate $\{\sigma_i\}$ of the system, the variable $N_K$ gives the number of instances of the FLS on the lattice. Fig.~\ref{fig:FLS}k shows an example how $N_K$ is determined for the L-shape FLS for a given configuration of the lattice: We visit every lattice site occupied by a particle of type A and check whether its eight neighbouring sites match the FLS of Fig.~\ref{fig:FLS}b-e, with the site marked by eyes at the location of the visited site. If so, the count $N_K$ gets increased by one. The total energy then reads
\begin{equation}
E(\{\sigma_i\}) = -J \sum_{\langle i,j\rangle} \sigma_i \sigma_j - K N_K,
\label{eq:energy}
\end{equation}
where the sum over nearest-neighbour pairs $\langle i,j\rangle$ conveys the Ising-type interaction.

The system is in contact with a heat bath at temperature $T$, corresponding to the inverse temperature $\beta=1/T$, with Boltzmann's constant set to unity. When we consider non-conserved particle numbers $N_A$ and $N_B$, in the presence of particle reservoirs with exchange chemical potential $\mu$, the grand canonical partition function is 
\begin{equation}
\mathcal{Z}_\mathrm{gc} = \sum_{\{\sigma_i = \pm 1\}} e^{-\beta E(\{\sigma_i\})  + \beta\mu(N_A - N_B)}.
\label{eq:grandZ}
\end{equation}
For conserved particle numbers, the canonical partition function is
\begin{equation}
\mathcal{Z}_\mathrm{can} = \sum_{\{\sigma_i = \pm 1|N_A,N_B\}} e^{-\beta E(\{\sigma_i\})},
\label{eq:canonZ}
\end{equation}
where the sum runs over all possible configurations with fixed particle numbers $N_A$ and~$N_B$.

\section{Ground states without number conservation}

\begin{figure}
\begin{indented}
\item[]
  \includegraphics[width=0.75\textwidth]{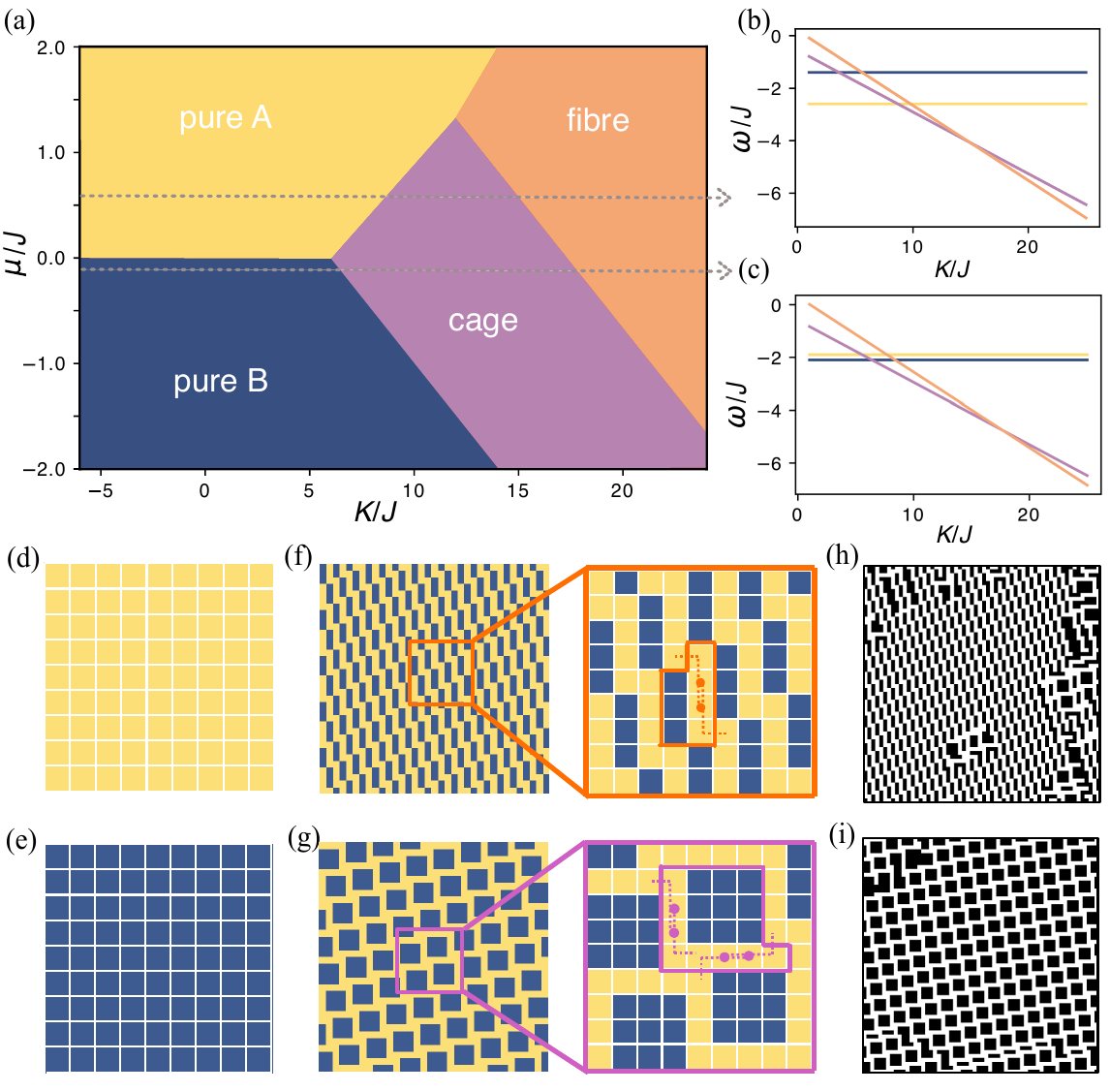}
   \caption{\label{fig:ground_states}% 
   Phase behaviour for the L shape as FLS, in the low temperature limit and without number conservation. Panel (a) shows the phase diagram of ground states as a function of the parameters $K$ and $\mu$. The ground states are identified through the minimisation of the grand potential per site $\omega$, shown in panels (b) and (c) for two selected values of the chemical potential, $\mu/J=0.6$ and $\mu/J=-0.1$ [indicated in panel (a) as dashed lines]. Line colours matching to the four different tiling patterns labeled in panel (a). The other panels show the lattice gas in the pure A (d), pure B (e), fibre (f,h), and cage (g,i) ground states. The ideal form is shown in panels (f) and (g), where the inset highlights the unit tile, with every instance of the FLS marked. We show representative snapshots of simulations at $T / J = 1.2$, with $K/J = 18$ (h) and $K/J = 8$ (i) for $\mu/J = 0$, which have lead to the identification of the two non-trivial ground states. 
   }
   \end{indented}
\end{figure}

To explore the system's behavior, we employ Monte Carlo simulations~\cite{frenkel2023understanding} at low temperatures using Metropolis sampling. While the total number of particles $N=N_A+N_B$ remains fixed, the numbers $N_A$ and $N_B$ change with every accepted Monte Carlo move, changing the local state of a lattice site. The results of these simulations allows us to identify ground states associated with different FLSs and different values of the parameters $K$ and $\mu$. While for small $K$, compared to the Ising interaction $J$, one expects the the uniform states of pure $A$ or pure $B$ to dominate, non-trivial ground states emerge for a sufficiently large $K$ of the FLS interaction. Typically, a value of $K$ several times bigger than $J$ is necessary in order to offset the energetic cost of having several unequal neighbouring sites within a FLS. For the case of the L-shape FLS, we find two ground states exhibiting a chiral periodic tiling, termed ``fibre'' (Fig.~\ref{fig:ground_states}h) and ``cage'' (Fig.~\ref{fig:ground_states}i). While the simulation results obtained through a slow decrease of the temperature still contain some defects, the ideal periodic tiling patterns and their unit tiles (Figs.~\ref{fig:ground_states}f,g) can be inferred.

A phase diagram, as shown in Fig.~\ref{fig:ground_states}a, can be obtained through the minimisation of the grand potential per lattice site, which in the zero-temperature limit is
\begin{equation}
    \omega\equiv [E-\mu(N_A-N_B)]/N.
    \label{eq:grandpot}
\end{equation}
We label each ideal tiling pattern by an index $i$ and calculate its grand potential density $\omega_i$ by evaluating Eq.~\eqref{eq:grandpot}. Generally, every tile is characterised by the number $n_{K,i}$ of instances of the FLS centred within the tile (but possibly spanning across neighbouring tiles) and the number $n_{J,i}$ associated with the Ising interaction. The latter counts the number of nearest-neighbour pairs with equal state, minus the number of such pairs with different state. To avoid over-counting, a pair contributes to $n_{J,i}$ if (and only if) the top or the left of the two neighbouring sites is located within the unit tile, the other neighbour may be located in one of its periodic images. For a unit tile containing $n_{A,i}$ sites of type A and $n_{B,i}$ sites of type $B$ (giving $n_i=n_{A,i}+n_{B,i}$ sites in total), we then obtain the overall grand potential per site 
\begin{equation}
    \omega_i= [-Jn_{J,i}-Kn_{K,i}-\mu(n_{A,i}-n_{B,i})]/n_i.
    \label{eq:grandpot2}
\end{equation}
This expression can also be written as $\omega_i=\varepsilon_i-\mu(c_{A,i}-c_{B,i})$, with the average energy per lattice site
\begin{equation}
    \varepsilon_i=(-Jn_{J,i}-Kn_{K,i})/n_i
    \label{eq:epsilondef}
\end{equation}
and the concentrations
\begin{equation}
    c_{A,i}=n_{A,i}/n_i\textup{ and }c_{B,i}=n_{B,i}/n_i
    \label{eq:concdef}
\end{equation}
specific for a tiling of type $i$.

For the system with the L-shape FLS, the coefficients characterising the four tiling patterns (``pure A'', ``pure B'', ``cage'' and ``fibre'') are given in Tab.~\ref{tab:co_ex}. They yield the following linear functions for the grand potential per site in the different phases:
\numparts
\begin{eqnarray}
    \omega_{A} &=& -\mu - 2J, \\
    \omega_{B} &=& \mu - 2J, \\
    \omega_{\mathrm{cage}} &=& (\mu -10J - 4K)/17, \\
    \omega_{\mathrm{fibre}} &=& (-\mu + 2J - 2K)/7.
\end{eqnarray}
\endnumparts
We choose $J$ as an energy scale, which leaves us to explore the phase behaviour as a function of the two scaled parameters $\mu/J$ and $K/J$.
The values of $\omega_i$ for the four different phases may intersect, as shown in Fig.~\ref{fig:ground_states}b.
The minimum of the grand potential over these four possibilities then yields the ground state for any given values of these parameters, leading to the phase diagram of Fig.~\ref{fig:ground_states}a. 

For small $K$ or large positive or negative $\mu$, the pure phases are always preferred. Increasing $K$ at fixed $\mu$ ultimately leads to the fibre phase. The cage phase is obtained for intermediary values of $K$ and sufficiently small $\mu$. The phase diagram has two triple points: One at $(K=6J,\mu=0)$ and one at $(K=12J,\mu=4J/3)$. The contact lines between any two phases in the phase diagram indicate singular values of $K$ and $\mu$ where two phases could, in principle, coexist in bulk. However, in a finite system supplied by a particle reservoir, the typically non-zero interfacial energy (see Sec.~\ref{sec:interface}) favours the existence of only a single phase. The coexistence lines for pure B and cage and for cage and fibre are parallel, such that there is never any direct transition from pure B to fibre.

\begin{table}[b]
\caption{\label{tab:co_ex}%
Tiling properties and energy per site $\varepsilon_i$ for the four ground states for the L-shape local FLS
}
\begin{indented}
\item[]\begin{tabular}{ccccc}
phase &pure A & pure B & cage &fibre\\
\hline
$n_{i}$& 1 & 1 & 17 & 7 \\
$n_{A,i}$& 1 & 0 & 8 & 4 \\
$n_{B,i}$& 0 & 1 & 9 & 3 \\
$n_{K,i}$& 0 & 0 & 4 & 2 \\
$n_{J,i}$& 2 & 2 & 10 & -2 \\
$c_{A,i}$& 1 & 0 & 8/17 & 4/7 \\
$\varepsilon_{i}$ & $-2J$ & $-2J$ & $(-10J - 4K)/17$ & $(2J - 2K)/7$\\
\end{tabular}
\end{indented}
\end{table}

\begin{figure}
\begin{indented}
\item[]
  \includegraphics[width=0.5\textwidth]{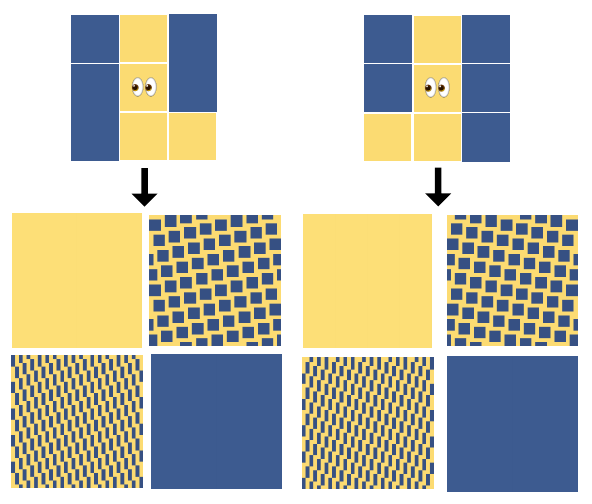}
   \caption{\label{fig:chiral}%
   The chirality of FLS can be reflected in the chirality of the tiling pattern of the ground state, as visible in the cage and fibre phases of the L- and J-type FLSs.
   } 
   \end{indented}
\end{figure}

Our analysis of the L-shape FLS shows how chirality present at the microscopic scale can be reflected on a larger scale. When the parameter $K$ is small, the phases pure A or pure B ensuing as ground states are mirror symmetric. When $K$ is sufficiently large to enter the cage or fibre phase, this symmetry is broken and the respective tiling patterns inherit the chirality of the FLS. As shown in Fig.~\ref{fig:chiral}, the cage and fibre patterns produced by the L- and J-shape FLSs cannot be mapped to one another by rotation, hence they have different chirality. We observe that while the cage pattern preserves the 4-fold rotational symmetry of the square lattice, the symmetry gets reduced to only 2-fold for the fibre pattern. Hence, there are two distinct fibre patterns, rotated to each other by $90^\circ$, there are two degenerate ground states for the parameters corresponding to the fibre phase, which have the same grand potential. A further degeneracy of the ground states is the translational offset of the tiling, for which there are $n_i$ possibilities. 

In the appendix, we show additional results for the examples of FLS shapes of Fig.~\ref{fig:FLS}. These confirm that we only see chiral tiling patterns for the two chiral FLS, namely the S and the h shape. Moreover, we see the trend that the phase behaviour becomes simpler for higher levels of symmetry of the underlying FLS. Only the L and h shapes, which have the same low symmetry, display two non-trivial tiling patterns as possible ground states. In contrast, we find only a single non-trivial tiling pattern for each of the shapes T, H and S. The latter is chiral, but compared to the L and h shapes it has an additional rotational symmetry.

\section{Phase coexistence for particle number conservation}

We now turn to the case where the numbers $N_A$ and $N_B$ of the two types of particles occupying the lattice are fixed. In a biological context, this case corresponds to a cell of a fixed volume with compartments formed of building blocks which are synthesised, degraded or exchanged with the extracellular environment on a much slower time scale than the one on which phase separation happens. We study how phase coexistence and phase separation in a two-component mixture are influenced by the presence of FLSs.

We impose a fixed fraction of lattice sites occupied by the two components ($c_A=N_A/N$ and $c_B=N_B/N$, where $c_A + c_B = 1$) and vary the strength $K$ of the FLS to study the coexistence behaviour.
For a system that is sufficiently large for the bulk energy to dominate over any interfacial contributions, the energy per site (i.e., the free energy per site at zero temperature) can be expressed as:
\begin{equation}
    \varepsilon = \sum_i \phi_i \varepsilon_i.
    \label{eq:energydensity}
\end{equation}
Here $\phi_i$ represents the fraction of area of the total lattice covered by phase $i$, and $\epsilon_i$ denotes its energy per site, satisfying $\sum_i \phi_i = 1$. The sum runs over all phases, e.g., for the L-shape FLS we have pure A ($i=1$), cage ($i=2$), fibre ($i=3$), or pure B ($i=4$). The energy per site of each of the phases is given by Eq.~\eqref{eq:epsilondef} and listed in Tab.~\ref{tab:co_ex} for the phases of the L-shape FLS.

\begin{figure*}
  \includegraphics[width=1\textwidth]{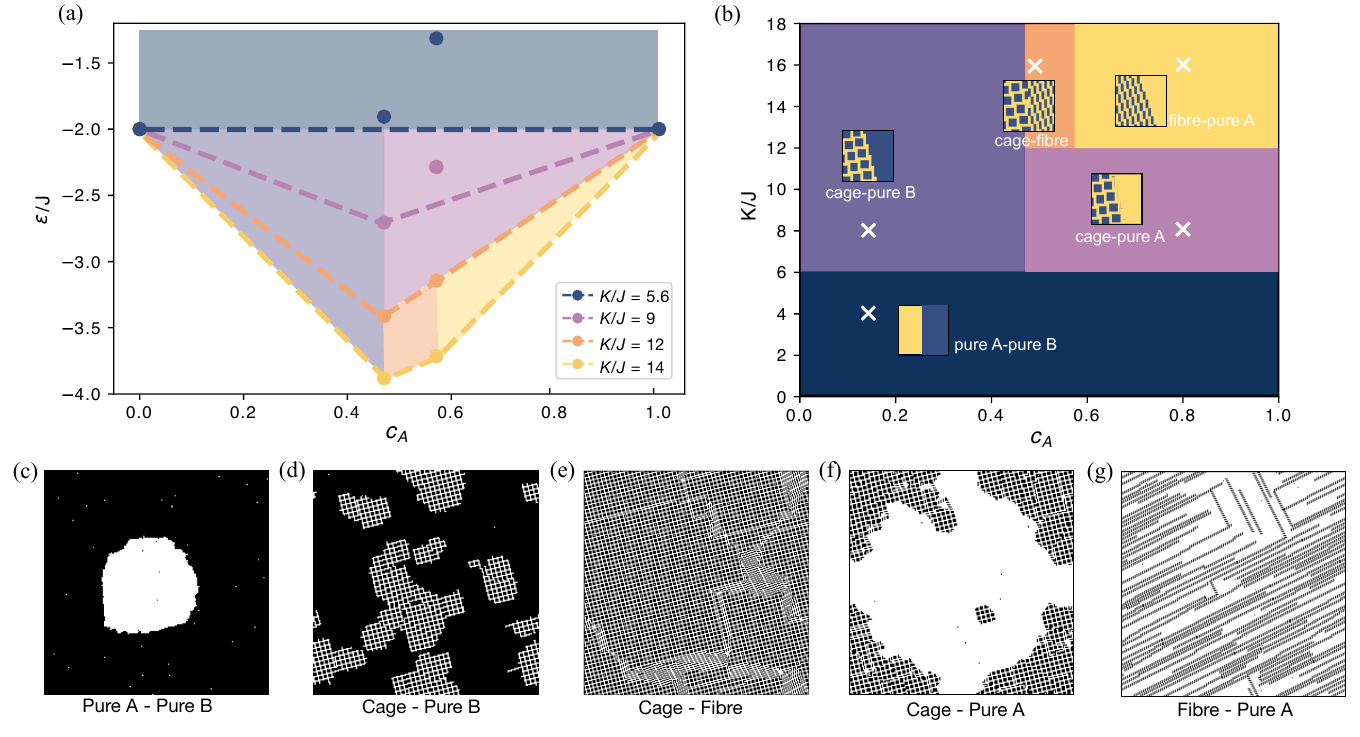}
   \caption{\label{fig:co_ex}% 
   (a) Energy per site of phase co-existence steady state. The filled areas represent the co-existence scenario with the same colours as in the phase diagram. The dots for each $K$ represent the energy densities for the overall fraction $c_A$ of A for the phases pure B, cage, fibre and pure A (from left to right). (b) Phase diagram of phase co-existence steady state at low temperature. White crosses correspond to the parameters of (c)-(g). (c)-(g) Final configurations of simulations at (c) $c_A=0.15, K/J=4$, (d) $c_A=0.15, K/J=8$, (e) $c_A=0.5, K/J=16$, (f) $c_A=0.8, K/J=8$, and (g) $c_A=0.8, K/J=16$. In each simulation, the temperature is continuously decreased from $T=1.6J$ to $T\simeq 1.07J$ (c,d,f) or from $T=5J$ to $T\simeq 1.9J$ (e,g). 
   }
\end{figure*}

To find the ground state, the energy of Eq.~\eqref{eq:energydensity} is to be minimised with respect to the fractions $\phi_i$ under the constraint
\begin{equation}
    c_A=\sum_i\phi_i c_{A,i}
\end{equation}
imposed by number conservation, where $c_{A,i}$ of Eq.~\eqref{eq:concdef} is the fraction of sites in state A in each of the phases. Moreover, the area fractions $\phi_i$ are constrained by $\sum_i\phi_i=1$ and $0\leq\phi_i\leq 1$. The minimisation procedure now follows the idea of a common tangent construction shown in Fig.~\ref{fig:co_ex}a, familiar from the physics of phase separation. The only difference is that the fraction $c_A$ for a single phase is not a continuous order parameter, but one that can only assume the discrete values $c_{A,i}$ associated with the different tiling patterns. Consider the values of $c_{A,i}$ and $\varepsilon_i$ as dots in a coordinate system. For the coexistence of two phases $i$ and $j$, covering the area fractions $\phi_j$ and $\phi_k$ with $\phi_i+\phi_j=1$, both $c_A$ and $\varepsilon$ change linearly with $\phi_i$. This traces out a coexistence line in the $c_A$-$\varepsilon$-plane that connects the dots corresponding to the two phases. For every given value of $c_A$, we can find the pair of coexisting phases from the coexistence line that minimises $\varepsilon$. 

The pair of coexisting phases depends on $c_A$ and the ratio $K/J$, as shown in the phase diagram of Fig.~\ref{fig:co_ex}b. When K is small, only pure A and pure B phases can coexist. However, when $K/J > 6$ the cage phase comes into play, coexisting with either pure~A or pure B, depending on $c_A$. The phases cage and pure B coexist for $c_A < c_{A,\mathrm{cage}}$, while the phases cage and pure A coexist for $c_A > c_{A,\mathrm{cage}}$ and $6 < K/J < 12$. When $K/J > 12$, the fibre phase starts to appear. For $c_{A,\mathrm{cage}} < c_A < c_{A,\mathrm{fibre}}$, fibre and cage phases coexist, while for $c_A > c_{A,\mathrm{fibre}}$, fibre and pure A phases coexist. It is impossible to form a stable fibre-pure B coexistence. The coexistence of three phases is restricted to special cases: Reducing both $\phi_i$ and $\phi_j$ for a pair of optimal, co-existing phases to allow for a third non-zero $\phi_k$ can never decrease $\varepsilon$, and it can only stay the same for values of $K/J$ corresponding to triple points (e.g., for $K/J=12$ in Fig.~\ref{fig:co_ex}a,b, where cage, fibre, and pure A can coexist).

We conduct Monte Carlo simulations at finite temperatures using Kawasaki dynamics \cite{kawasaki1972kinetics} to complement our theory. This involves randomly selecting two lattice sites and attempting to swap the particles occupying them, with acceptance or rejection determined by the Boltzmann weights associated with the change in energy. This procedure allows us to reach a steady state of the system. Starting from a random initial state, the ground state at zero temperature can be approximated through a slow reduction of temperature (``simulating annealing'') \cite{frenkel2023understanding}. The simulation results for the L-shape FLS (Fig.~\ref{fig:co_ex}c-g) confirm that the phases identified above indeed coexist.

The simulations also show the typical fluctuations occurring at small, non-zero temperature. For $K/J$ well below $6$, as in Fig.~\ref{fig:co_ex}c, fluctuations are similar to those of the regular Ising model, showing a few particles of the opposite type dispersed into the bulk phases. For the phases involving tiling patterns, fluctuations often show up in the form of defect lines, where grains of mismatching offsets of the patterns touch. Moreover, we observe for the cages phase in Fig.~\ref{fig:co_ex}e that the central lattice site of a cage can flip to state A. Such a flip leaves the number of FLSs in the local tile unaffected. Its energetic cost of $8J$ can be overcome by fluctuations at sufficiently high temperature and the freed B-type particle can then be bound in additional FLSs created elsewhere. 

\section{Structure of interfaces between co-existing phases}
\label{sec:interface}

\begin{figure}
  \includegraphics[width=1\textwidth]{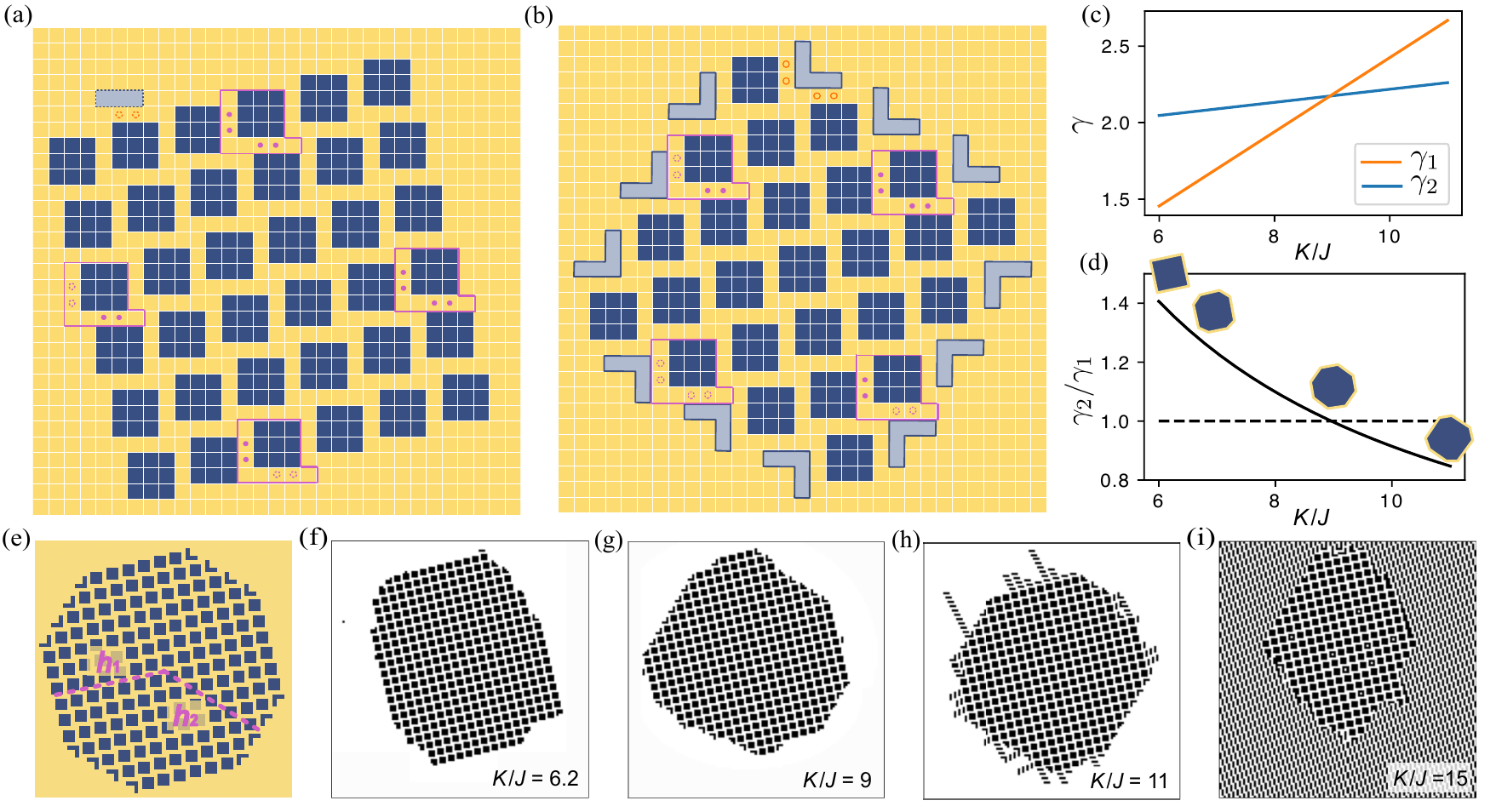}
   \caption{Interface properties of a condensate formed by the cage phase coexisting with pure A. 
   (a) Square shaped condensate with interface of type 1. The pink frame marks a tile unit, and pink dots denote the centres of L-shape FLSs within that unit. The blue shaded bar is a proposed interface decoration which is not energetically favoured, despite the two additional FLSs it produces (dash-lined orange circles).  (b) Square shaped condensates with interface of type 2. The pink frame marks the tile unit, and pink dots denote L-shape FLSs. The blue shaded `V's are an energetically favoured decoration, producing four additional FLSs (orange open circles). (c) Interfacial energy of interfaces of type 1 and type 2 as a function of the relative strength of the FLS interaction. (d) Ratio $\gamma_2/\gamma_1$, which determines the indicated crystalline shape of condensates in Wulff construction. (e) An octagonal condensate with interfaces of both types. The heights of the interfaces above the geometric centre are denoted as $h_1$ and $h_2$, respectively. (f-h) Final configurations of simulations of the cage-pure A coexistence at low temperature with $K/J = 6.2$ (f), $K/J = 9$ (g), and $K/J = 11$ (h). (i) Final configuration of a simulation of the cage-fibre coexistence at $K/J=15$. Each simulation starts from a circular cage condensate and reduces the temperature from  $T=1.8$ to $T\simeq1.1$ (f-h) or from $T=4$ to $T\simeq2.4$ (i).}
   \label{fig:interfaces}% 
\end{figure}

We now focus on the structure and energy of interfaces between coexisting phases of the lattice gas. In particular for finite size systems, the interfacial energy presents an important contribution to the overall energy, besides the bulk energies we have derived so far. It ultimately determines the shape and orientation of a condensate formed in the limit of low temperature. 

Consider the coexistence of two phases, denoted 1 and 2, in a setting with number conservation. An area $A_1$ is attributed to phase $1$ and an area $A_2$ to phase $2$. Throughout this section, we use the lattice spacing as a unit of length. Therefore the area of a phase also corresponds to the total number of lattice sites it occupies. Our goal is to express the energy of a phase separated system as
\begin{equation}
E=A_1\varepsilon_1+A_2\varepsilon_2+\sum_\nu\gamma_\nu\ell_\nu.
\label{eq:crystalenergy}
\end{equation}
The first two terms represent the bulk energies, while the summation introduces the interfacial energies of different types of interfaces, labelled by $\nu$. These types differ by the orientation of the interface with respect to the lattice axes or by interfacial patterns, as we discuss below. The line density of interfacial energy is denoted as $\gamma_\nu$ and the length of interfaces as $\ell_\nu$. We anticipate that when the energy is minimal, the regions of coexisiting phases have the geometry of polygons, with a finite number of different interface types. 

At interfaces, we can observe patterns that are not part of the tiling pattern of either of the two phases. These give interfaces a certain ``thickness'' of a few lattice spacings, up to which the areas $A_1$ and $A_2$ are \textit{a~priori} ill-defined. Even in the limit of large areas, these ambiguities enter Eq.~\eqref{eq:crystalenergy} to the same order as the interfacial terms and are therefore non-negligible. We aim to fix the definitions of $A_1$, $A_2$, and, accordingly, $\gamma_\nu$ such that they remain invariant for any change of shape of the coexisting regions that conserve the overall numbers $N_A$ and $N_B$. Useful, general definitions of the areas $A_1$ and $A_2$ are obtained by requiring $A_1+A_2=N$ and $A_1c_{A,1}+A_2c_{A,2}=N_A$, yielding 
\begin{equation}
    A_1=\frac{N_A-Nc_{A,2}}{c_{A,1}-c_{A,2}},\qquad A_2=\frac{N_A-Nc_{A,1}}{c_{A,2}-c_{A,1}}.
    \label{eq:Adef}
\end{equation}
They are invariant because they depend only on conserved quantities. The thickness of interfaces also makes the definition of the $\ell_\nu$ ambiguous. We define them as the Euklidian length of straight lines drawn along the interfaces, measured in units of the lattice constant. Small errors in $\ell_\nu$ of the order of a few lattice constants contribute to the energy of Eq.~\eqref{eq:crystalenergy} to the same order as the energy associated with corners between straight edges. For sufficiently large condensates both can be safely neglected.

Once it is ensured that $A_1$ and $A_2$ are conserved numbers, the minimisation of the energy~\eqref{eq:crystalenergy} can be performed through the so-called Wulff construction \cite{wulff1901xxv}. It requires that the distances $h_\nu$ of interfaces above the geometric centre of a crystal (as indicated in the example of Fig.~\ref{fig:interfaces}e) are proportional to $\gamma_\nu$. These heights ultimately determine the polygonal shape of the crystal.

For concreteness, we now turn to the co-existence of the phases cage (phase 1) and pure A (phase 2) in the model of L-shape FLS. From the bulk thermodynamics described above, we determine that this coexistence is possible for $6\leq K/J\leq 12$ (see also Fig.~\ref{fig:co_ex}b). We start by constructing a square-shaped condensate, as shown in Fig.~\ref{fig:interfaces}a. 
While this square shape may not minimise the energy of the system, it allows us to identify the energy of a single type of interface, before proceeding to polygons that have different types of interface. We arrange $m\times m$ unit tiles of the cage phase in the form of a square lattice, forming a large square surrounded by the pure A phase. Due to the shape of the unit tiles, the square is tilted with respect to the lattice axes, such that the upper interface has a slope of $1/4$, as shown in Fig.~\ref{fig:interfaces}a. We label this interface and, equivalently, the three other interfaces rotated in steps of $90^\circ$ as interfaces of type $\nu=1$. The area covered by tiles of the cage pattern is $A_1=17m^2$ (recall the tile size $n_\mathrm{cage}=17$) and the remaining area covered by the pure A phase is $A_2=N-A_1$. The total number of lattice sites in state A is $N_A=A_1c_{A,1}+A_2c_{A,2}$ (with $c_{A,1}=8/17$ and $c_{A,2}=1$) and the total length of the interface surrounding the square is $\ell=4m\sqrt{4^2+1^2}$ [reflecting the $(4,1)$ translational symmetry]. The energy of this configuration is
\begin{equation}
    E=-2JN+24Jm^2-(4m^2-4m) K=A_1\varepsilon_1+A_2\varepsilon_2+4mK,
    \label{eq:energy14}
\end{equation}
which can be seen as follows. Starting from a pure A phase covering the whole lattice (energy $-2JN$), every cage-tile places a $3\times 3$ square of B into a neighbourhood of A, which causes 12 new pairs of unequal neighbouring sites and increases the energy by $24J$. This holds for every such tile, both in the bulk and at the interface (this observation is specific for the cage-pure A coexistence, in general there could be a difference in the contribution to the Ising-type energy for bulk and surface tiles). However, for the energy of the FLS, there are differences between the bulk and surface contributions. Most tiles contain the centres of four L's and thus contribute the energy $-4K$, as marked by dots for a tile of the upper surface shown in Fig.~\ref{fig:interfaces}a. Yet, for tiles forming part of the left or lower edges of the cage-phase, the lack of neighbouring tiles allows them to host only two centres of L's. Comparing the overall energy Eq.~\eqref{eq:energy14} to \eqref{eq:crystalenergy} finally yields the interfacial energy 
\begin{equation}
    \gamma_1=K/\sqrt{17}
    \label{eq:gamma1}
\end{equation}
for this type of interface between cage and pure A. Due to symmetry, this interfacial energy is the same for upper, lower, left, and right interfaces, despite apparent differences related to the ambiguous definition of the unit tile.

Beyond the simple type of interface we have discussed so far, there may be more complex interfaces that feature ``decorations'' that are not native to either of the two unit tiles.
For the coexistence of the cage and pure A phase, typical interface decorations observed in simulations at finite temperature are bars of three B sites above an interface of slope $1/4$ and V-shapes of five B sites above an interface of slope $-3/5$, as shown in shaded blue in Fig.~\ref{fig:interfaces}a and b.
Adding such decorations typically produces a surplus of either A or B sites on the surface. In a number conserved ensemble, as considered here, these particles need to be taken from the bulk phase, which affects the overall energy. 
A natural way to assign areas to the two coexisting phases is to take the areas $A_1^0$ and $A_2^0$ covered by the respective tiles prior to adding any decorations. That way, we can identify the surplus $n_{A,\nu}^+$ of A particles per length of the interface of type $\nu$ through
\begin{equation}
    N_A=A_1^0c_{A,1}+A_2^0 c_{A,2}+\sum_\nu\ell_\nu n_{A,\nu}^+.
\end{equation}
However, the areas $A_1^0$ and $A_2^0$ differ slightly from the areas $A_1$, $A_2$ defined in Eq.~\eqref{eq:Adef}. Simple algebra shows that the bulk energies associated with these two different definitions of area differ as
\begin{equation}
    A_1^0\varepsilon_1+A_2^0\varepsilon_2=A_1\varepsilon_1+A_2\varepsilon_2-\sum_\nu n_{A,\nu}^+\ell_\nu\frac{\varepsilon_1-\varepsilon_2}{c_{A,1}-c_{A,2}}.
\end{equation}
When changing from one definition of area to the other, the last term gets attributed to the interfacial energy in Eq.~\eqref{eq:crystalenergy}. This term equals $2n_{A,\nu}^+\ell_\nu\mu$, where $\mu$ is the chemical potential for which the grand canonical potential \eqref{eq:grandpot2} is equal for the two coexisting phases. For sufficiently large systems, the bulk phase acts as a particle reservoir with this exchange chemical potential, from which the particles required for the decoration can be recruited. For pure A-cage coexistence we have $\mu=2K/9-4J/3$

Consider the bar decoration of Fig.~\ref{fig:interfaces}a for interfaces of slope $1/4$. It allows two additional FLSs to appear (marked by two dashed centres of L's), but costs an Ising energy of $16J$ (flipping 8 pairs of neighbouring sites from $-J$ to $+J$) and requires 3 A-type particles being replaced by B-type particles, in exchange with the bulk phase. Hence we get the overall energy change of this decoration per length $\ell$ of the decorated interface $\Delta E/\ell=(-2K+16J+6\mu)/\sqrt{17}=(8J-2K/3)/\sqrt{17}$. This change becomes negative for $K>12J$. However, $K=12J$ marks the end of the pure A-cage coexistence, such that this interface decoration is never thermodynamically favourable for $T\to 0$. Nonetheless, this decoration presents a metastable state which is often visited in simulations at finite temperature and for $K/J$ close to 12. Often these bars appear stacked on top of each other, forming long protrusions as visible in Fig.~\ref{fig:interfaces}h.

\renewcommand{\arraystretch}{1.2}
\begin{table*}
\caption{\label{tab:int_sample}
Examples for interfacial patterns and their energy for coexistent phases of the L-shape FLS.
}
\footnotesize
\begin{tabular}{c|c|c|c|c}
 & pattern & slope & $K/J$ & $\gamma$ \\\hline
pure A - pure B & \includegraphics[width=1.5cm, height=1.5cm]{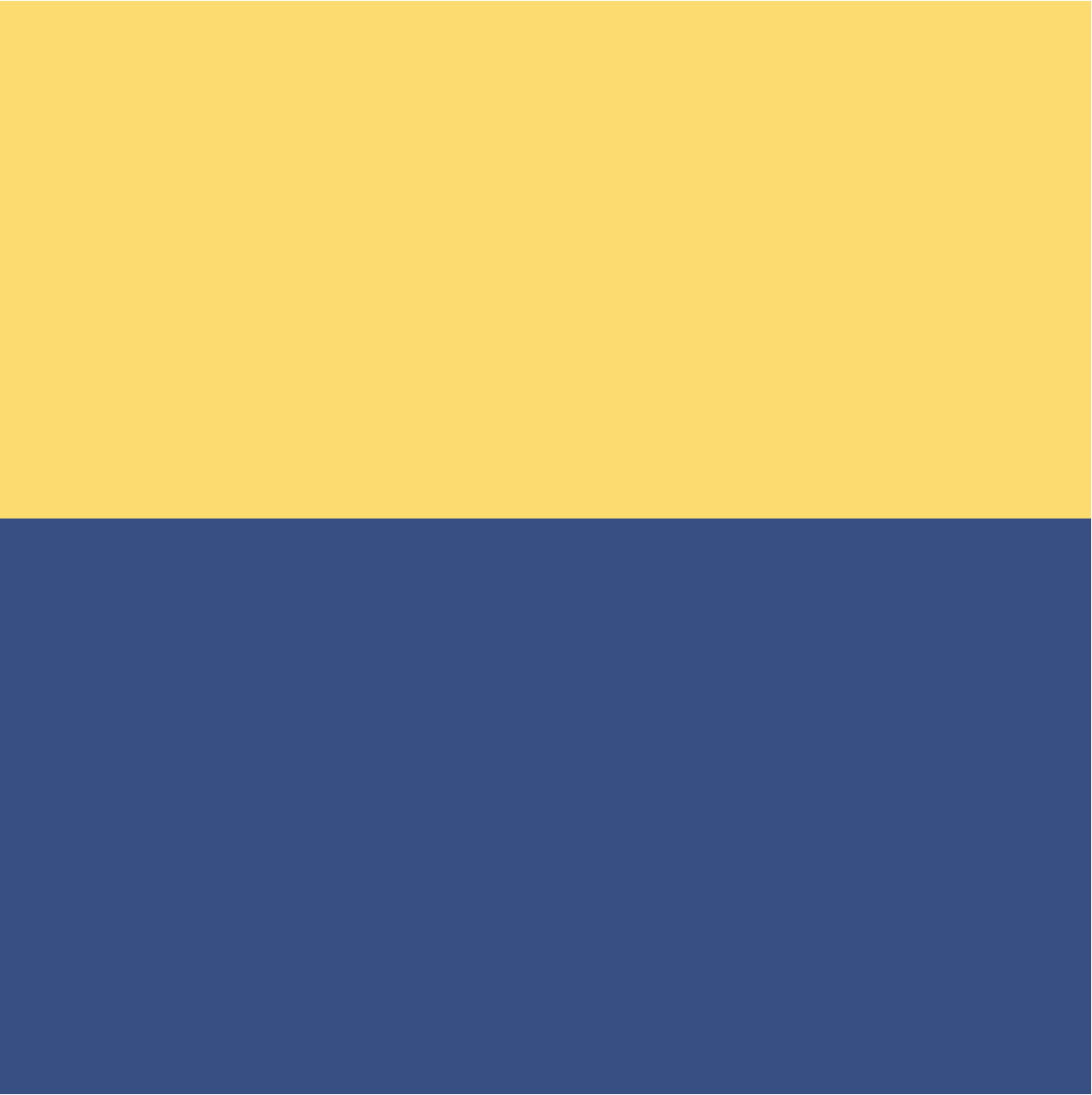} & $0$ & $ 0 \leq K/J \leq 6$ & $2J$ \\\hline
Cage - pure B & \includegraphics[width=1.5cm, height=1.5cm]{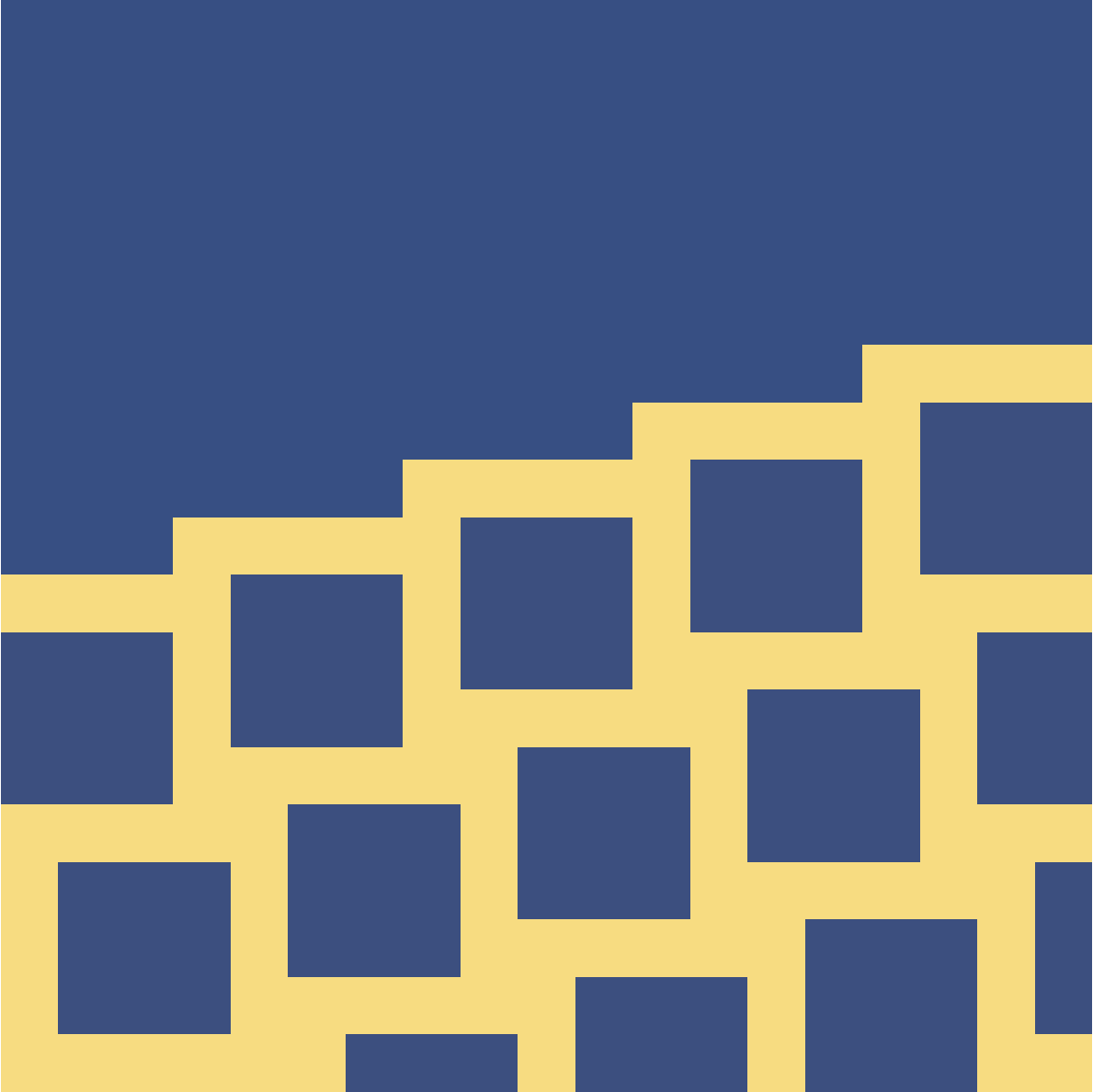} & $1/4$ & $K/J > 6$ & $(K/2-J)/\sqrt{17}$ \\
 & \includegraphics[width=1.5cm, height=1.5cm]{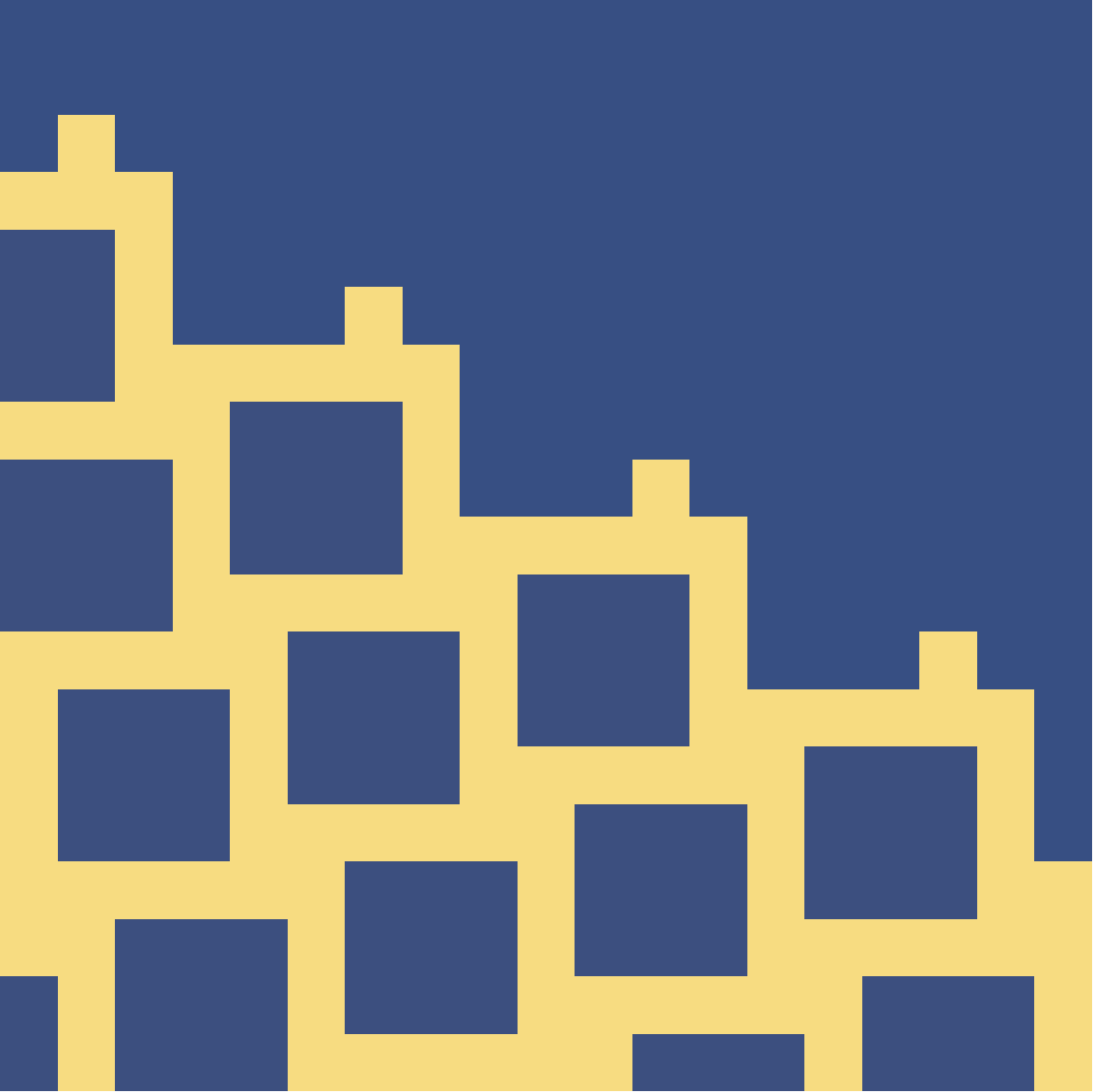} & $-3/5$ & $K/J > 14$ & $(K/2+5J)/\sqrt{34}$ \\\hline
Cage - pure A & \includegraphics[width=1.5cm, height=1.5cm]{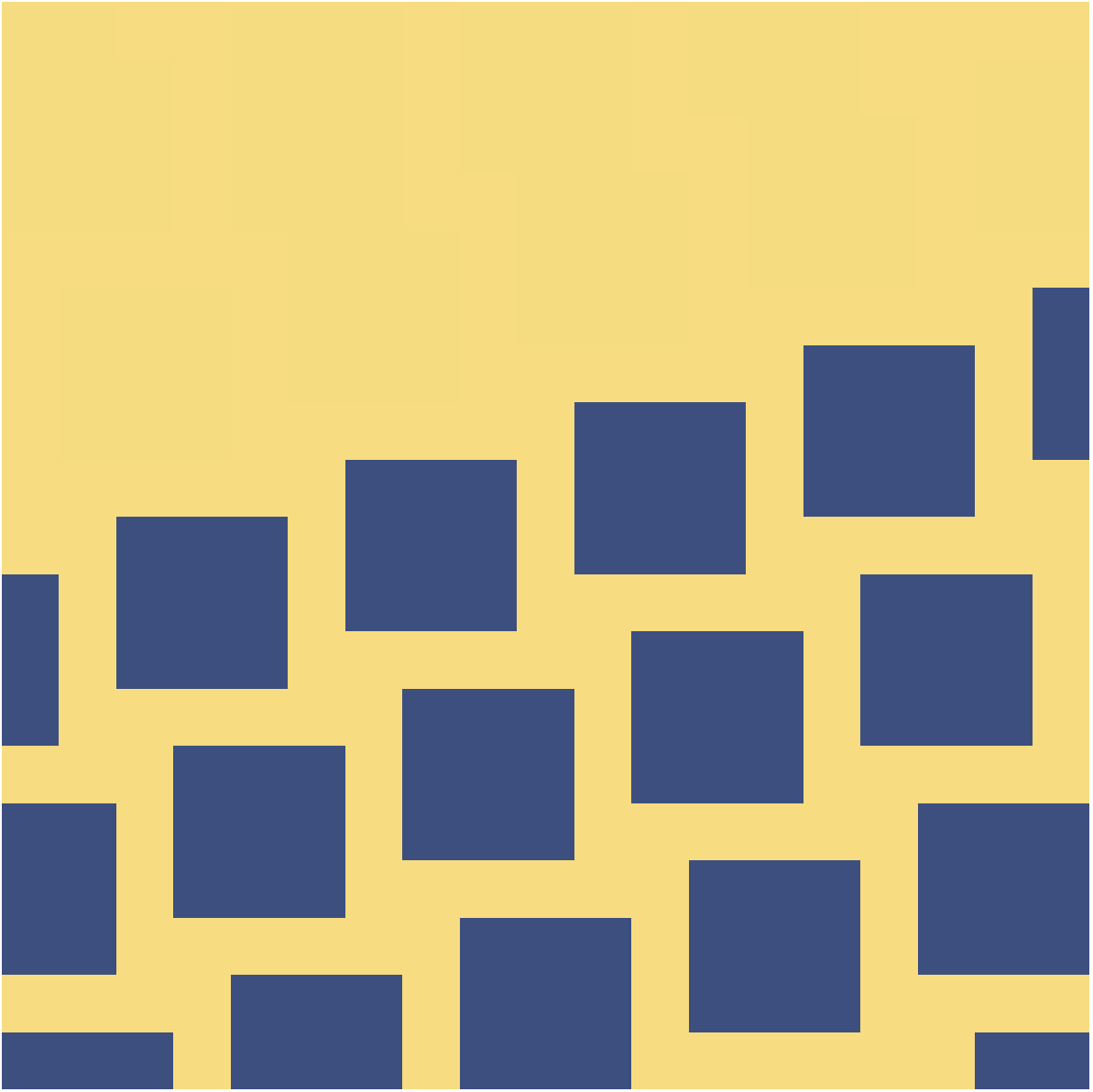} & $1/4$ & $6 \leq K/J < 12$ & $K/\sqrt{17}$ \\
 & \includegraphics[width=1.5cm, height=1.5cm]{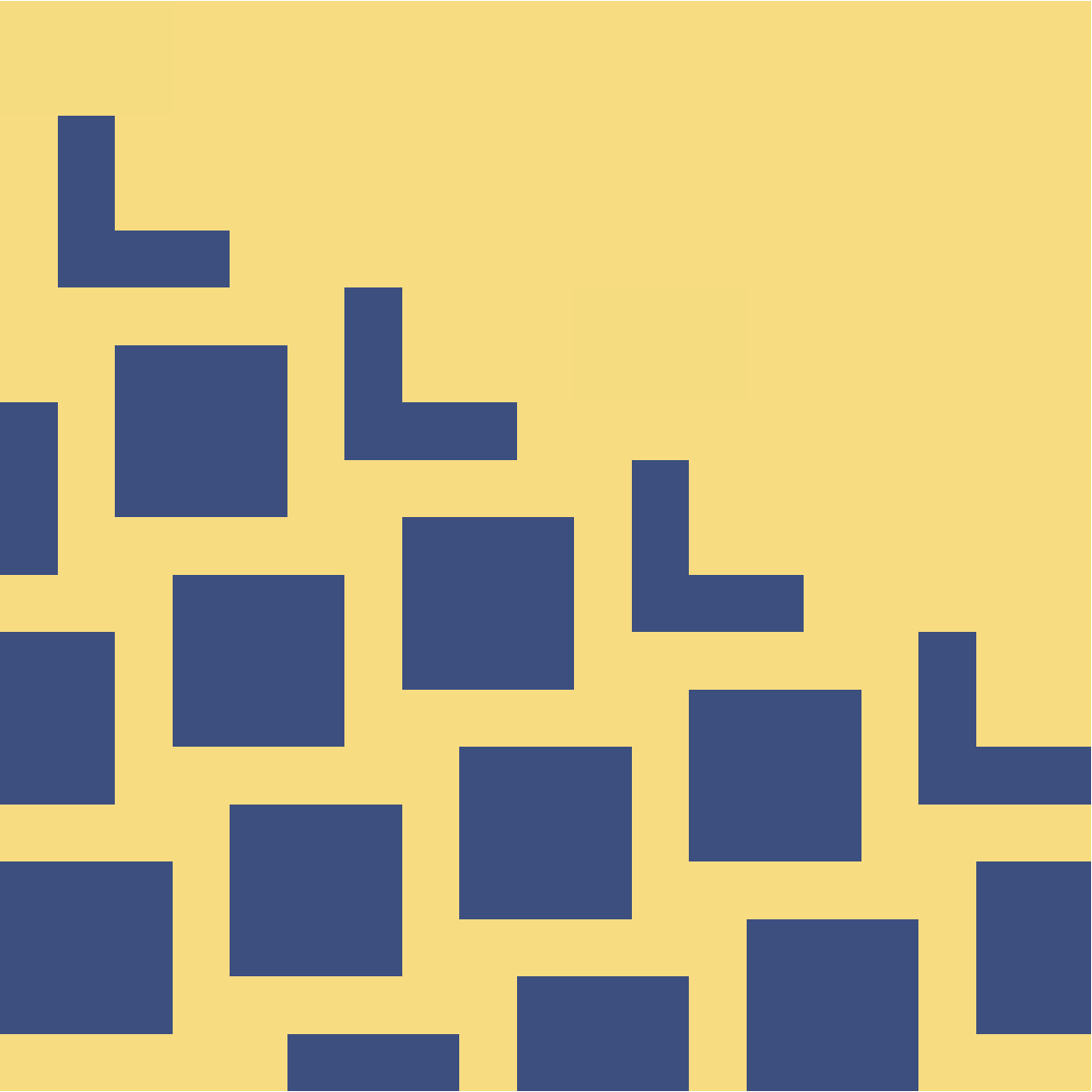} & $-3/5$ & $6 < K/J < 12$ & $\displaystyle\sqrt{\frac{2}{17}}\left(\frac{K}{9}+\frac{16J}{3}\right)$ \\\hline
Cage - Fibre & \includegraphics[width=1.5cm, height=1.5cm]{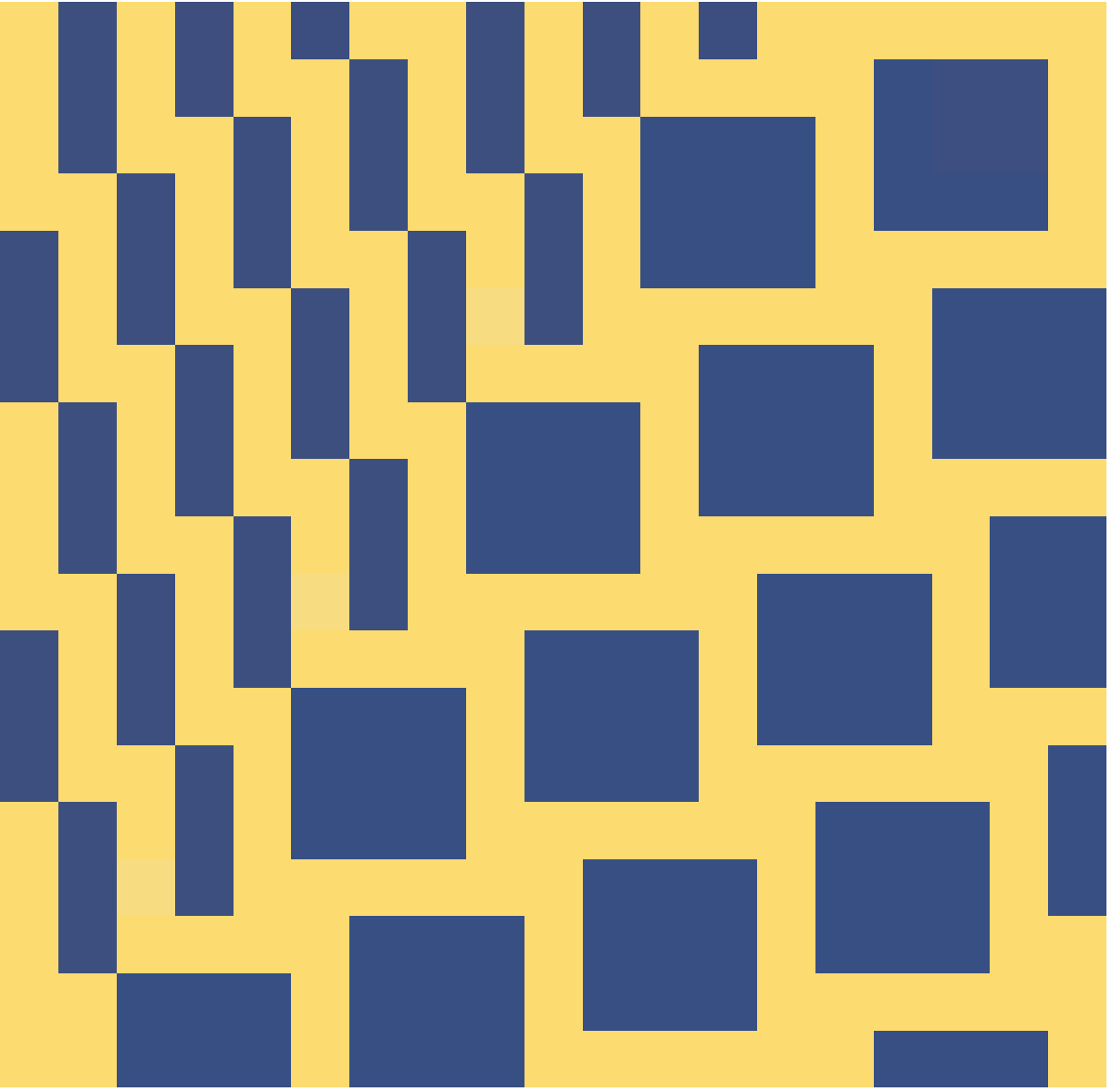} & $5/3$ & $12 \leq K/J$ & $(K-8J)/\sqrt{34}$ \\
 & \includegraphics[width=1.5cm, height=1.5cm]{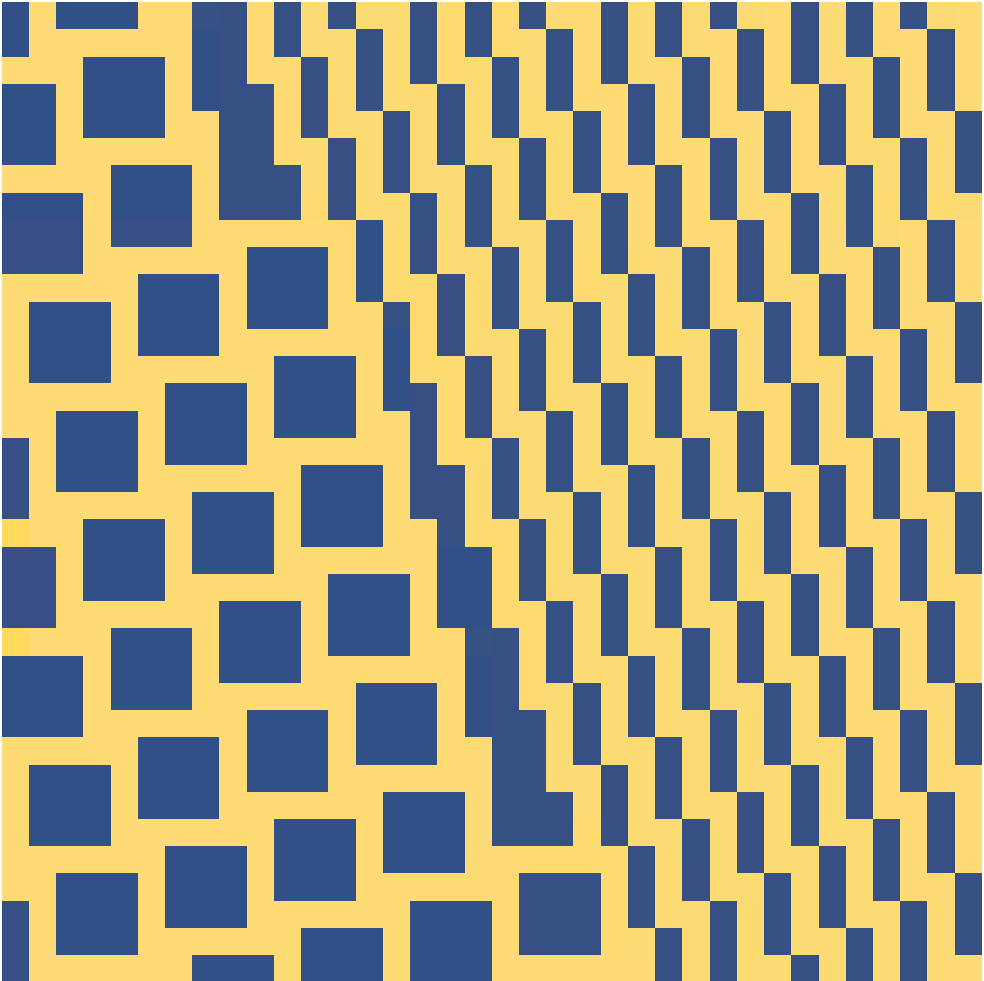} & $-23/10$ & $12 \leq K/J$ & $(80J/3+16K)/\sqrt{629}$ \\\hline
pure A - Fibre & \includegraphics[width=1.5cm, height=1.5cm]{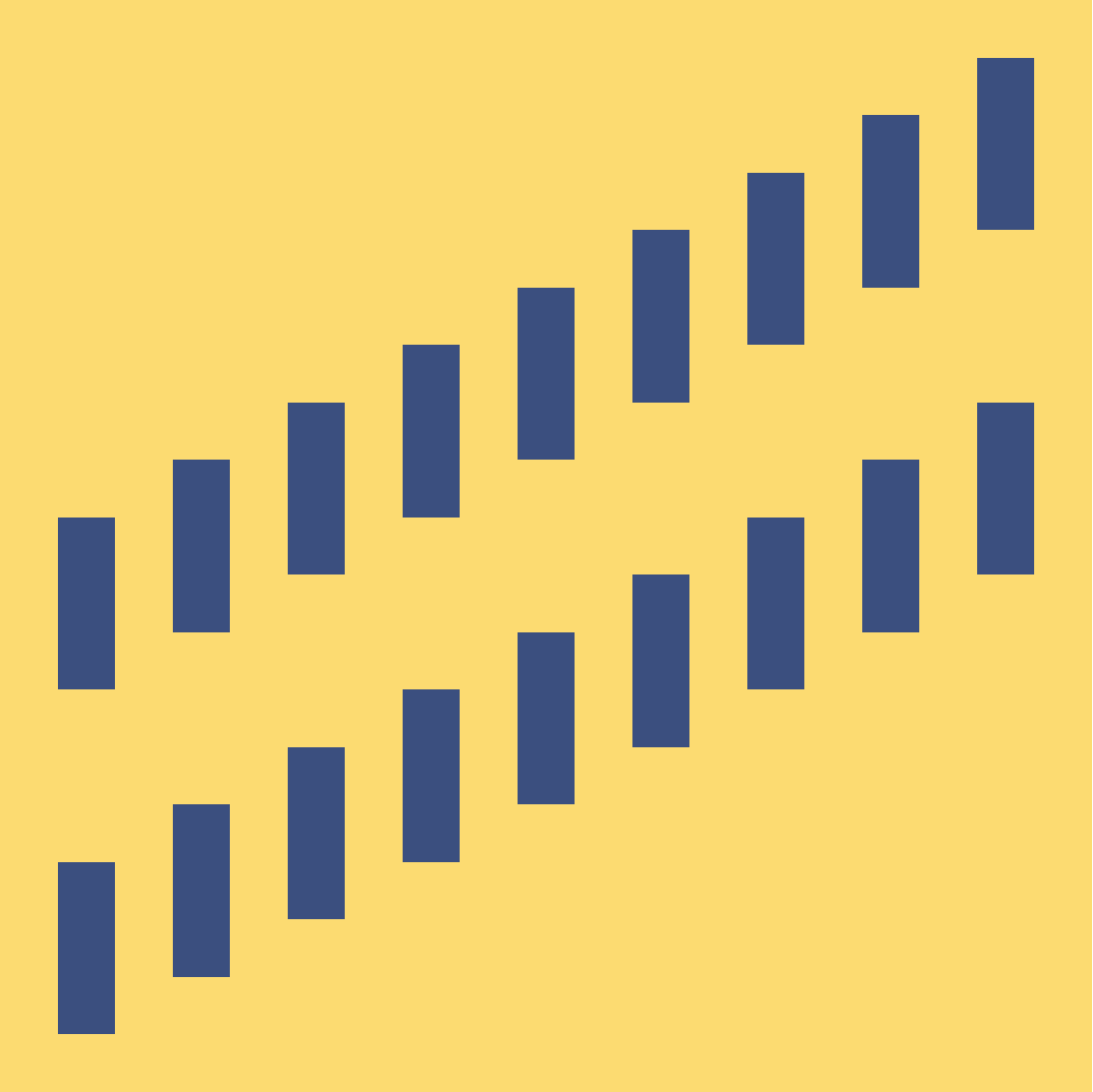} & $1/2$ & $12 \leq K/J$ & $0$ \\
\end{tabular}
\end{table*}

Another interface showing up in simulations of the pure A-cage coexistence is labelled type $\nu=2$. It has slope $-3/5$ (for instance), cutting diagonally across the square lattice of the cage tiling, as shown in Fig.~\ref{fig:interfaces}b. Following a similar argument as for the interface of type 1 above, we find that without decorations, the interfacial energy is $\gamma=2K/\sqrt{3^2+5^2}$. However, this energy can be further reduced by adding V-shaped decorations formed of five B particles. They produce four more L's per surface tile at the cost of an additional Ising energy of $24J$, leading to 
\begin{equation}
    \gamma_{2}=\frac{1}{\sqrt{34}}(2K-4K+24J+10\mu)=\sqrt{\frac{2}{17}}\left(\frac{K}{9}+\frac{16J}{3}\right).
    \label{eq:gamma2}
\end{equation}
This interfacial energy is less than that of the undecorated interface of the same slope for all values of $K$ and $J$ where pure A and cage coexist. Hence, the decorations are favoured for this kind of interface.

Putting the results for the interfacial energy of the pure A-cage coexistence together, we can determine the shape of condensates of minimal energy through the Wulff construction. The interfacial energies for the two different types of interface are equal for $K\approx 8.9J$, as shown in Fig.~\ref{fig:interfaces}d.  In that case, the shape of minimal energy is a regular octagon. Otherwise, the height $h_2$ of the $-3/5$ slope above the crystal centre decreases with respect to the height $h_1$ of the $1/4$ slope, according to the decreasing ratio of the respective interfacial energies shown in Fig.~\ref{fig:interfaces}e. Given the angle $45^\circ$ between the two types of interfaces, square shaped condensates become favourable when the ratio is above $\sqrt{2}$ or below $1/\sqrt{2}$. This boundary is reached at $K=6J$, where the optimal shape would be a simple square tilted by the slope $1/4$ with respect to the lattice axes. However, this value also presents the lower end of the coexistence region of cage and pure A. On the other end, octagonal condensates (with increasingly dominant $-3/5$ slopes) prevail up to $K=12J$, where the fibre phase takes over.

Tab.~\ref{tab:int_sample} shows the patterns and properties of a few other interfaces encountered for L-shape FLSs. For the coexistence of cage and pure B, we find again interfaces of slopes $1/4$ and $-3/5$. Here, the slope $1/4$ is energetically preferred so much for small $K$, such that crystals remain square shaped up to $K=14J$. For larger $K$ one obtains octagonal crystals. As shown in the phase diagram of Fig.~\ref{fig:co_ex}, cage and pure B can coexist for arbitrarily large $K$. The ratio of the interfacial energy of the slope $-3/5$ to that of the slope $1/4$ then approaches $1/\sqrt{2}$ from above, such that crystals remain octagonal for any large $K$ and approach a square shape with slope $-3/5$ only for $K\to\infty$.

For the coexistence of pure A and fibre, encountered for sufficiently large $K$ and large overall concentration of A, we find that an interface of slope $1/2$ (for upward oriented fibres) comes at zero energetic cost, while interfaces of any other slope have positive interfacial energy. This explains why simulations (such as in Fig.~\ref{fig:co_ex}g) show bars arranged as many long and thin ascending strands. For the fibre phase with fibres oriented horizontally, these strands descend at slope $-2$.

Because the tiling patterns of the fibre and the cage phase have different periodicities, the coexistence between the two can lead to rather complex interface structures. While it is prohibitive to analyse all possible interface patterns, we list in Tab.~\ref{tab:int_sample} the properties of two types of interfaces that are predominant in simulations of this type of coexistence, as shown in Fig.~\ref{fig:interfaces}i. An interface of slope $5/3$ is the only one where the two tiling patterns can be matched in a simple fashion. For a fibre phase that is everywhere oriented in a single direction, this type of interface comes only in two orientations, rotated by $180^\circ$ to each other. Hence, another type of interface is needed to form a compact condensate. This interface has a zig-zag structure of overall slope $-23/10$, with the pure B phase acting as a buffer between the two mismatching tiling patterns. The resulting condensates of the cage phase have the structure of a parallelogram.  \par
The simulation results of Fig.~\ref{fig:interfaces}f-i largely confirm the shapes of
condensates we predict theoretically for the limit of zero temperature. In addition, they
provide insight into the fluctuations that are expected for temperatures that are small
compared to the interaction strength $K$ but yet non-zero. Here, we have chosen the temperature small enough and simulation times long enough to eliminate almost all defects in the bulk phases. Defects at the interface between phases include the above mentioned metastable decorations, as well as additional corners that interrupt the energetically favoured straight interfaces. Such defects can be created through the occasional ``evaporation'' of a particle from the interface (as captured on the left side in the snapshot shown in Fig.~\ref{fig:interfaces}f), which then diffuses through the bulk phase and gets re-adsorbed at a random position on the interface. The resulting non-ideal shapes of condensates are entropically favoured at non-zero temperature.

\section{Discussion and Conclusions}

The motivation for this work has been to study broken chiral symmetry in a minimal equilibrium model. For this purpose, we have extended the model of an Ising-type lattice gas by a chiral FLS with next-nearest-neighbour interactions.
We have shown that such a model can display a rich phase behaviour, depending on the strength $K$ of the FLS interaction compared to the regular Ising interaction $J$ and on other parameters.

We have illustrated our findings for the example of a FLS that prefers the formation of an L-shape by the particles of any local environment.
We have focused on ground states at zero temperature and started with the grand canonical ensemble. Varying $K$ and the chemical potential, we have identified four possible ground states. Besides the two pure, achiral ground states of the traditional Ising model, two chiral ground states with cage and fibre tiling patterns emerge for sufficiently strong $K$. Being tilted with respect to the lattice axes, these tiling patterns reflect the microscopic breaking of chiral symmetry at the macroscopic scale.

With particle number conservation, we analytically find five different scenarios of phase coexistence in the low-temperature limit. Condensates are composed of a patch of one of the previously identified ground states, suspended in another one. Thermodynamics predicts which phases can coexist for given $K$ and fixed overall particle numbers of each type.

For finite systems, the interfaces between such phases plays an important role. Here, the structure may differ from the bulk. We have developed a methods to calculate interfacial energies of these structures.
With these interfacial energies at hand, one can determine the polygonal, crystalline shape of condensates using the Wulff construction. We find that for condensates of the cage or fibre phase, the chiral nature of the model is revealed not only in the tiling pattern of the bulk, but also in the characteristic slopes of interfaces with respect to the lattice axes. For the case of the cage-fibre coexistence, condensates assume the for of a parallelogram, which can be identified as chiral even without reference to the lattice axes.

While we have picked the L-shape FLS for illustration purposes, our methodology applies to any kind of FLSs, such as the ones shown in  Fig.~\ref{fig:FLS}, whether they are chiral or achiral. FLSs will always lead to more ground states than pure A and pure B, with the chirality of the tiling and interface patterns determined by the chirality of the chosen FLS (see Appendix). While we have limited our examples to $3\times 3$ FLSs on square lattices, none of our mathematical results depend on this geometry. In particular, they are also applicable to hexagonal lattices, as studied in Refs.~\cite{ronceray2011variety,ronceray2012geometry,ronceray2013influence,ronceray2014multiple}.

We have focused on analytical calculations in the limit of zero temperature. The insights gained from these calculations explain well the behaviour observed in simulations at small, finite temperatures. Further analytical progress in this regime could be made by considering possible excitations from the ground state and defect lines between tiling patterns with mismatching offset. At much higher temperature, one can expect critical behaviour, which can be studied using renormalisation group techniques. In a disordered, high temperature phase condensates can no longer form. Nonetheless it may be fruitful to explore the role of FLSs and chirality in an expansion for small $\beta$.

Our work presents a minimal approach to breaking chiral symmetry at equilibrium in binary mixtures. It could provide a starting point towards understanding self-assembly in more complex systems in higher dimensions and beyond square lattices, including, for instance the formation of condensates of chiral polymers in a continuum space \cite{furthauer2012active}.  
While our work shows that already the stationary, equilibrium behaviour of such chiral systems can be non-trivial, recent research shows that passive systems with a chiral microscopic structure display an odd response to being driven out of equilibrium by external forces \cite{zhao2022odd,lier2022passive,huang2023odd}.  Future work may build on the results presented here to study the dynamics of a chiral lattice gas in response to external perturbations or thermal fluctuations, and in complex environments~\cite{reichhardt2017depinning}.

Beyond the passive model considered here, it would be interesting to consider active variants, where the detailed balance of the update moves is broken. 
Such a chiral active lattice gas might display behaviour known from particle based and coontinuum models such as unique transport properties ~\cite{yang2021topologically,lier2023lift}, oscillatory phase behaviour ~\cite{liu2020oscillating, tan2022odd} and odd rheological response \cite{avron1998odd,banerjee2017odd, scheibner2020odd, banerjee2021active,markovich2021odd}.

\appendix

\section{Additional favoured local structures}

In the main text, we have chosen to exemplify our general formalism using the L-shape FLS, which is particularly interesting due to its chirality and rich phase behaviour. For comparison, we show here some results for the shapes shown in Fig.~\ref{fig:FLS}. 

\begin{table}[b]
\caption{\label{tab:h_and_s}%
Tiling properties and energy per site $\varepsilon_i$ for the new ground states for different FLS.
}
\small
\begin{tabular}{ccccccc}
phase & stripe (I) &cage (S) &cage (T) &checker-plate (h) & Pythagorean (h)& dot (H) \\
\hline
$n_{i}$& 2 & 10 & 9 & 13 & 13 & 4\\
$n_{A,i}$& 1 & 6 & 5 & 9 & 8 & 3\\
$n_{B,i}$& 1 & 4 & 4 & 4 & 5 & 1\\
$n_{K,i}$& 1 & 2 & 4 & 4 & 4 & 2\\
$n_{J,i}$& 0 & 4 & 2 & 2 & 2 & 0\\
$c_{A,i}$& 1/2 & 3/5 & 5/9 & 9/13 & 8/13 &3/4\\
$\omega_{i}$ & $\frac{-K}{2}$ & $\frac{-\mu-2J-K}{5}$ & $\frac{-\mu-2J-4K}{9}$ & $\frac{-5\mu-2J-4K}{13}$ &  $\frac{-3\mu-2J-4K}{13}$ & $\frac{-\mu-K}{2}$\\
\end{tabular}
\end{table}

\begin{figure}
  \includegraphics[width=1\textwidth]{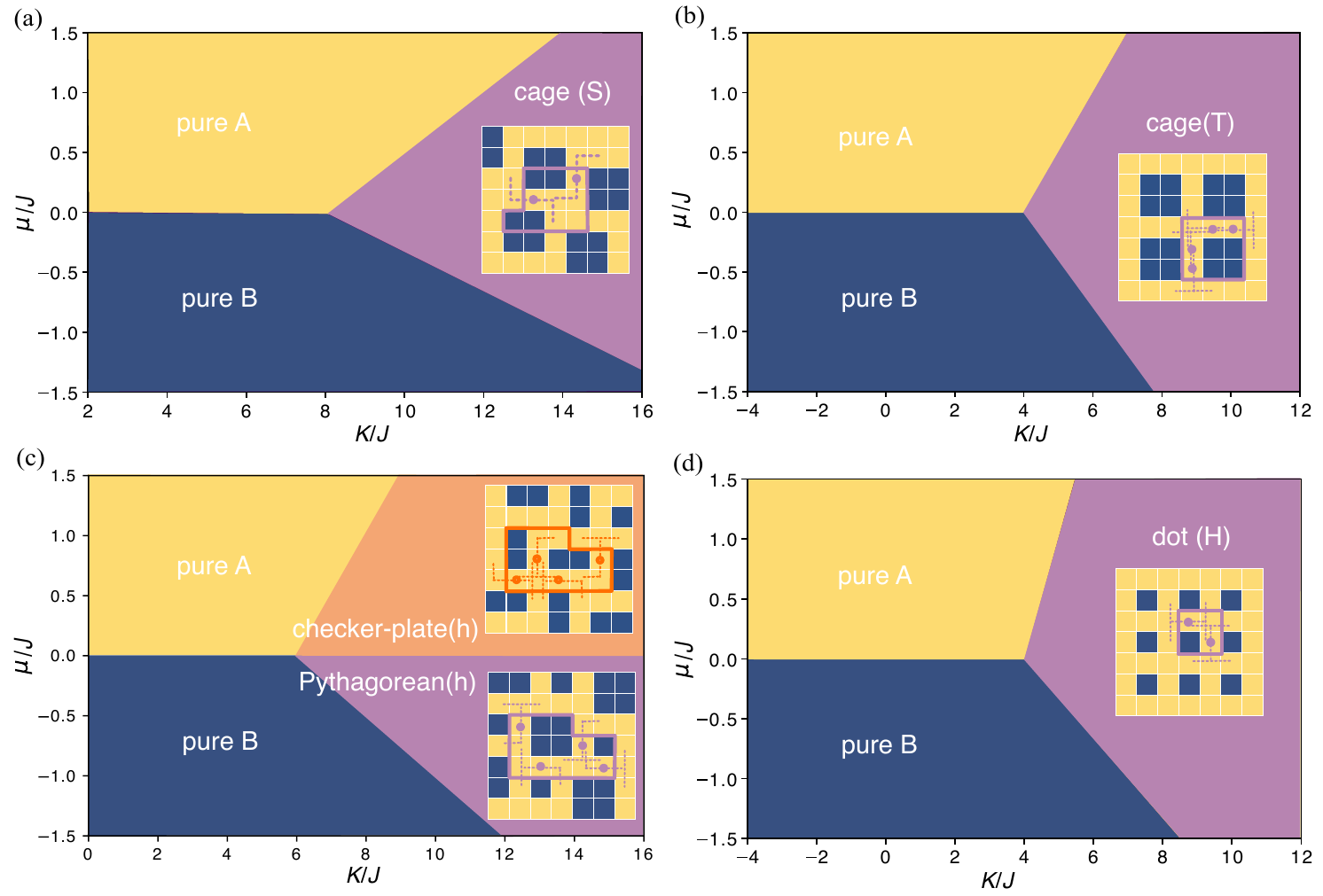}
   \caption{The phase
diagram of ground states, for the S shape (a), T shape (b), h shape (c)and H shape (d) as FLS, as a function of the parameters $K$ and $\mu$.}
\label{fig:phase_diagram_HS}
\end{figure}

In the grand canonical ensemble, only the h- and the L-shape model display four distinct phases, while in our other examples (I, S, T, H) there are only three thermodynamically stable phases. In addition to the already established pure A and pure B phases, we find for sufficiently large $K$ a variety of new tiling patterns. The simplest of these is a ``stripe phase'' of the I shape FLS, consisting of alternating stripes of A and B, each being one lattice side wide. Moreover, there is a phase we call ``dot'' for the H-shape FLS and a cage pattern for the T-shape FLS, both forming regular square tilings that are parallel to the underlying lattice, see the insets of the first column of Fig.~\ref{fig:phase_diagram_HS}. For the S-shape FLS, there is a tilted cage pattern (Fig.~\ref{fig:phase_diagram_HS}b), which differs from the one of the L-shape FLS, forming only $2\times 2$ squares of B particles. For the h-shape FLS, we name the two new phases shown in Fig.~\ref{fig:phase_diagram_HS}d ``checker-plate'' and ``pythagorean''. The latter corresponds to a Pythagorean tiling with two sizes of square tiles formed of particles of type B, with particles of type A forming gaps between them. Table~\ref{tab:h_and_s} lists the properties of the new phases, analogously to Tab.~\ref{tab:co_ex}. The properties of the pure A and pure B phase remain the same as in Tab.~\ref{tab:co_ex}. Through the $K$ and $\mu$-dependent minimisation of the grand potential $\omega$ over the three phases for each model, we obtain the phase diagrams shown in Fig.~\ref{fig:phase_diagram_HS}.

\begin{figure}
  \includegraphics[width=1\textwidth]{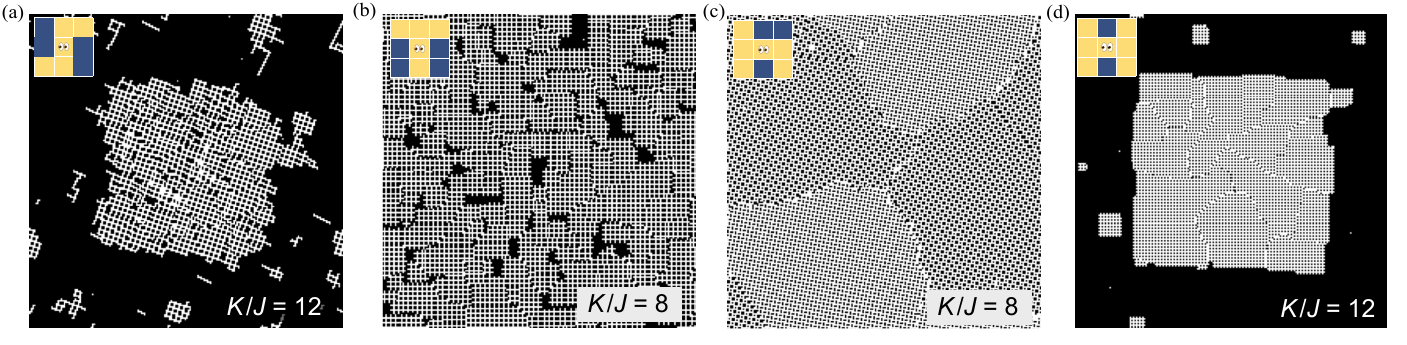}
   \caption{Low temperature simulation snapshots of condensate formation for S-shape, T-shape, h-shape and H-shape FLSs, as indicated by the insets. Parameters: (a),(d) $K/J=12$ and (b),(c) $K/J=8$.}
   \label{fig:hs_dplt}
\end{figure}

Simulations at low temperature with conservation of particle numbers show the formation of condensates, as shown in Fig.~\ref{fig:hs_dplt}. The achiral nature of the H-shape and T-shape models is revealed by the vertical and horizontal interfaces. 
The coexistence of the cage and pure B phase for the S-shape FLS is similar to the L-shape one. Interfaces of slope $1/3$ are preferred and reflect the chirality of the FLS. As a common, yet metastable interface decoration one can observe staircase-like protusions. The two non-trivial phases of h-shape model both have the same tile unit, which allows for interfaces of the slopes $2/3$ and $-1/5$.

\section*{References}

\bibliographystyle{iopart-num.bst}
%\nocite{*}
\bibliography{literature.bib}

\end{document}